\documentclass[aps,prl,twocolumn,showpacs,superscriptaddress,nobibnotes]{revtex4-1}

\usepackage{graphicx}
\usepackage[utf8]{inputenc}
\usepackage{siunitx}
\usepackage{amsmath}
\usepackage{gensymb}
\usepackage{amssymb}
\usepackage{enumerate}
\usepackage{color}
\usepackage{hyperref}
\usepackage[american]{babel}
\usepackage{etoolbox}
\usepackage{floatrow}
\usepackage{nicefrac}

\patchcmd{\thebibliography}{\section*{\refname}}{}{}{}

\newcommand{\aegis}{AE$\bar{\hbox{g}}$IS}

\newcommand{\pos}{$\mathrm{e^+}$}

\newcommand{\orb}[3]{$#1^{#2}\text{#3}$}
\newcommand{\ttP}[0]{\orb{3}{3}{P}}

\newcommand{\otS}[0]{\orb{1}{3}{S}}

\newcommand{\tP}[0]{\orb{2}{3}{P}}

\DeclareSIUnit\inch{in.}
\DeclareSIUnit\division{div}

\usepackage{lineno}

\begin{document}

\preprint{APS/123-QED}
\title{Positronium laser cooling via the \otS{}--\tP{} transition with a broadband laser pulse}
\date{\today}
\pacs{36.10.Dr, 32.80.Pj, 78.70.Bj}

\newcommand{\corresponding}[1]{\altaffiliation{Corresponding author: #1}}

\newcommand{\UofLatvia}[0]{\affiliation{University of Latvia, Department of Physics Raina boulevard 19, LV-1586, Riga, Latvia}}
\newcommand{\prague}[0]{\affiliation{Czech Technical University, Prague, Brehova 7, 11519~Prague~1, Czech Republic}}
\newcommand{\IAEP}[0]{\affiliation{Institute of Experimental and Applied Physics, Czech Technical University in Prague, Husova 240/5, 110 00, Prague 1, Czech Republic}}
\newcommand{\brescia}[0]{\affiliation{Department of Mechanical and Industrial Engineering, University of Brescia, via Branze 38, 25123~Brescia, Italy}}
\newcommand{\infnpv}[0]{\affiliation{INFN Pavia, via Bassi 6, 27100~Pavia, Italy}}
\newcommand{\tn}[0]{\affiliation{Department of Physics, University of Trento, via Sommarive 14, 38123~Povo, Trento, Italy}}
\newcommand{\infntn}[0]{\affiliation{TIFPA/INFN Trento, via Sommarive 14, 38123~Povo, Trento, Italy}}
\newcommand{\oslo}[0]{\affiliation{Department of Physics, University of Oslo, Sem Sælandsvei 24, 0371~Oslo, Norway}}
\newcommand{\infnmi}[0]{\affiliation{INFN Milano, via Celoria 16, 20133~Milano, Italy}}
\newcommand{\torun}[0]{\affiliation{Institute of Physics, Faculty of Physics, Astronomy, and Informatics, Nicolaus Copernicus University in Torun, Grudziadzka 5, 87-100 Torun, Poland}}
\newcommand{\polimi}[0]{\affiliation{Department of Aerospace Science and Technology, Politecnico di Milano, via La Masa 34, 20156~Milano, Italy}}
\newcommand{\cern}[0]{\affiliation{Physics Department, CERN, 1211~Geneva~23, Switzerland}}
\newcommand{\hamburg}[0]{\affiliation{Institute for Experimental Physics, Universit\"{a}t Hamburg, 22607 Hamburg, Germany}}
\newcommand{\warsaw}[0]{\affiliation{Warsaw University of Technology, Faculty of Physics, ul. Koszykowa 75, 00-662, Warsaw, Poland}}
\newcommand{\warsawel}[0]{\affiliation{Warsaw University of Technology, Faculty of Electronics and Information Technology, ul. Nowowiejska 15/19, 00-665 Warsaw, Poland}}
\newcommand{\vienna}[0]{\affiliation{Stefan Meyer Institute for Subatomic Physics, Austrian Academy of Sciences, Boltzmanngasse 3, 1090~Vienna, Austria}}
\newcommand{\infnpd}[0]{\affiliation{INFN Padova, via Marzolo 8, 35131~Padova, Italy}}
\newcommand{\RRI}[0]{\affiliation{Raman Research Institute, C. V. Raman Avenue, Sadashivanagar, Bangalore 560080, India}}
\newcommand{\pv}[0]{\affiliation{Department of Physics, University of Pavia, via Bassi 6, 27100~Pavia, Italy}}
\newcommand{\bs}[0]{\affiliation{Department of Civil, Environmental, Architectural Engineering and Mathematics, University of Brescia, via Branze 43, 25123~Brescia, Italy}}
\newcommand{\PAS}[0]{\affiliation{Institute of Physics, Polish Academy of Sciences, Aleja Lotnikow 32/46, PL-02668 Warsaw, Poland}}
\newcommand{\UoL}[0]{\affiliation{Department of Physics, University of Liverpool, Liverpool L69 3BX, UK}}
\newcommand{\CID}[0]{\affiliation{The Cockcroft Institute, Daresbury, Warrington WA4 4AD, UK}}
\newcommand{\NTNU}[0]{\affiliation{Department of Physics, NTNU, Norwegian University of Science and Technology, Trondheim, Norway}}
\newcommand{\leipzig}[0]{\affiliation{Felix Bloch Institute for Solid State Physics, Universit\"{a}t Leipzig, 04103 Leipzig, Germany}}
\newcommand{\mi}[0]{\affiliation{Department of Physics ``Aldo Pontremoli'', University of Milano, via Celoria 16, 20133~Milano, Italy}}
\newcommand{\afflac}[0]{\affiliation{Universit\'e Paris-Saclay, CNRS, Laboratoire Aim\'e Cotton, 91405, Orsay, France.}}

\author{L.~T.~Gl\"{o}ggler}
\cern

\author{N.~Gusakova}
\cern
\NTNU

\author{B.~Rien\"{a}cker}
\corresponding{brienaec@liverpool.ac.uk}
\UoL

\author{A.~Camper}
\corresponding{antoine.camper@fys.uio.no}
\oslo

\author{R.~Caravita}
\corresponding{ruggero.caravita@cern.ch}
\infntn

\author{S.~Huck}
\cern
\hamburg

\author{M.~Volponi}
\cern
\tn
\infntn

\author{T.~Wolz}
\cern

\author{L.~Penasa}
\tn
\infntn

\author{V.~Krumins}
\cern
\UofLatvia

\author{F.~Gustafsson}
\cern

\author{M. Auzins}
\UofLatvia

\author{B.~Bergmann}
\IAEP

\author{P.~Burian}
\IAEP

\author{R.~S.~Brusa}
\tn
\infntn

\author{F.~Castelli}
\infnmi
\mi

\author{R.~Ciury\l{}o}
\torun

\author{D. Comparat}
\afflac   

\author{G.~Consolati}
\infnmi
\polimi

\author{M.~Doser}
\cern

\author{\L.~Graczykowski}
\warsaw

\author{M.~Grosbart}
\cern

\author{F.~Guatieri}
\tn
\infntn

\author{S.~Haider}
\cern

\author{M.~A.~Janik}
\warsaw

\author{G.~Kasprowicz}
\warsawel

\author{G.~Khatri}
\cern

\author{\L.~K\l osowski}
\torun

\author{G.~Kornakov}
\warsaw

\author{L.~Lappo}
\warsaw

\author{A.~Linek}
\torun

\author{J.~Malamant}
\oslo

\author{S.~Mariazzi}
\tn
\infntn



\author{V.~Petracek}
\prague

\author{M.~Piwi\'nski}
\torun

\author{S.~Pospisil}
\IAEP

\author{L.~Povolo}
\tn
\infntn

\author{F.~Prelz}
\infnmi

\author{S.~A.~Rangwala}
\RRI

\author{T.~Rauschendorfer}
\cern
\leipzig

\author{B.~S.~Rawat}
\UoL
\CID

\author{V.~Rodin}
\UoL

\author{O.~M.~R{\o}hne}
\oslo

\author{H.~Sandaker}
\oslo

\author{P.~Smolyanskiy}
\IAEP

\author{T.~Sowi\'nski}
\PAS

\author{D.~Tefelski}
\warsaw

\author{T.~Vafeiadis}
\cern

\author{C.~P.~Welsch}
\UoL
\CID

\author{M.~Zawada}
\torun

\author{J.~Zielinski}
\warsaw

\author{N.~Zurlo}
\infnpv
\bs

\collaboration{The \aegis{} collaboration}
\noaffiliation{}

\begin{abstract}
We report on laser cooling of a large fraction of positronium (Ps) in free-flight by strongly saturating the \otS{}--\tP{} transition with a broadband, long-pulsed \SI{243}{\nano\meter} alexandrite laser. The ground state Ps cloud is produced in a magnetic and electric field-free environment.
We observe two different laser-induced effects. The first effect is an increase in the number of atoms in the ground state after the time Ps has spent in the long-lived \tP{} states. The second effect is the one-dimensional Doppler cooling of Ps, reducing the cloud's temperature from \SI{380\pm20}{\kelvin} to \SI{170\pm20}{\kelvin}. We demonstrate a \SI{58\pm9}{\percent} increase in the coldest fraction of the Ps ensemble.
\end{abstract}

\maketitle{}

\begin{figure}[htbp]
	\centering
	\includegraphics[width=\textwidth]{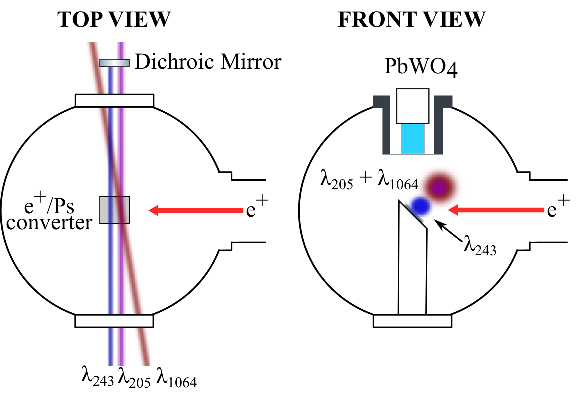}
	\caption{Simplified experimental scheme of the vacuum chamber where Ps is emitted from the \pos{}/Ps converter upon \pos{} implantation. The cooling laser ($\lambda_{243}$) and the two probing lasers pulses ($\lambda_{205}$ and $\lambda_{1064}$) enter and exit the chamber through view-ports. The $\mathrm{PbWO_4}$ crystal is represented in cyan, with the white rectangle above depicting the photo-multiplier tube. The photoionization laser ($\lambda_{1064}$) is injected under a slight angle, to avoid laser-induced damage to the dichroic mirror.}
	\label{fig:exp_scheme}
\end{figure}

Positronium (Ps), discovered in 1951, is the lightest known atomic system, consisting only of an electron and a positron (\pos{}) \cite{deutsch_ps:51}. Ps has been extensively studied for its exotic properties as a purely leptonic matter-antimatter system. So far, experiments researching Ps have relied on formation processes that result in clouds with a large velocity distribution, in the order of several $10^4\,$\SI{}{\meter\per\second} \cite{cassidy_silica:10, aegis_velo:20, aegis_morpho:21}. This, for instance, has been limiting the precision of spectroscopy studies due to the large Doppler broadening of the transition lines \cite{fee_1s2s_prl:93, fee_1s2s_pra:93}. 
The idea of using Ps laser cooling to narrow the velocity distribution dates back to 1988 \cite{Liang1988}, following the first demonstration of laser cooling on neutral atoms by just a few years \cite{RevModPhys.70.721}. Despite significant efforts \cite{kumita_pscooling:02}, Ps laser cooling has not been experimentally achieved yet. 
A whole range of new fundamental experiments would become feasible with a sufficient amount of cold Ps~\cite{cassidy_coldPs:08, cassidy_review:18}.  
These include 1S--2S precision spectroscopy at the \SI{100}{\kilo\hertz} level, which will enable testing bound state QED  at the  $\alpha^7 \, m_{e^+}$ order~\cite{cassidy_review:22}, measuring the $m_{e^+}/m_{e^-}$ mass ratio with unprecedented accuracy~\cite{Amsler2008_ParticleDataGroup}, and testing the Equivalence Principle (EP) with a purely leptonic system by looking at the transition red-shift around the Sun's orbit~\cite{Karshenboim_2016}. 
Testing the EP with atomic systems consisting of antimatter is the primary goal of the \aegis{} collaboration. \aegis{} builds on the availability of cold Ps for efficient antihydrogen ($\mathrm{\overline{H}}$) production through the charge exchange reaction $\mathrm{Ps^* + \overline{p} \rightarrow \overline{H}^* + e^-}$ between cold antiprotons ($\mathrm{\overline{p}}$) and Ps excited to Rydberg states (Ps*)~\cite{charlton:90}, first demonstrated experimentally in Ref.~\cite{aegis_pulsedhbar:21}. The charge exchange cross-section can be significantly increased by reducing the temperature of the Ps* cloud \cite{krasnicky_pra:16}. Moreover, forming a Ps Bose-Einstein condensate (BEC)~\cite{PhysRevB.49.454,cassidy_coldPs:08} will allow studying stimulated annihilation, producing coherent light in the gamma radiation range~\cite{varenna:10, avetissian_beclaser:14}. This objective, together with precision spectroscopy, is currently being pursued by the UTokyo group~\cite{Yamada2021}, which is actively developing Ps laser cooling with a different methodology employing a chirped laser pulse~\cite{Shu2016}.
\\
\indent Here, we report on the first experimental demonstration of Ps laser cooling by strongly saturating the \otS{}--\tP{} transition for \SI{70}{\nano\second}, employing an alexandrite-based laser system developed specifically to meet the requirements of Ps laser cooling (high intensity, large bandwidth, long pulse duration). The velocity distributions with and without laser cooling were obtained by Doppler sensitive two-photon resonant ionization along the \otS{}--\ttP{} transition~\cite{aegis_velo:20}. 
\\
Ps is produced by implanting a bunch of positrons with \SI{3.3}{\kilo\electronvolt} kinetic energy into a nanochannel-array etched into a silicon substrate~\cite{mariazzi_prb:10, aegis_morpho:21}. This \pos{}/Ps converter is mounted at an angle of \SI{45}{\degree} with respect to the \pos{} beam axis as shown in Fig.~\ref{fig:exp_scheme}. About \SI{30}{\percent} of the implanted positrons are re-emitted as Ps. An electrostatic buncher~\cite{aegis_nimb:15} with fast potential switching-off~\cite{aegis_meta2:18} complemented by a mu-metal shield were developed to conduct these experiments in a magnetic and electric field-free environment. The measured residual magnetic field was below \SI{1}{\milli\tesla} in the Ps production area. This development is important in view of Ps laser cooling, as it was shown that in the intermediate magnetic field range, the saturation of the \otS{}--\tP{} transition leads to a fast annihilation of the atoms due to singlet-triplet state mixing in the excited state manifold~\cite{Ziock1990_opticalsaturation}. This effect, called ``magnetic~quenching", prevents efficient Ps laser cooling~\cite{zimmer_cooling:21} and is a strong incentive to work either in the Paschen-Bach regime~\cite{cassidy_paschen:11} or, as we do, in a magnetic field-free environment. 

The Ps \otS{}--\tP{} transition is driven by the third harmonic of a Q-switched alexandrite laser~\cite{alex}, as proposed in Ref.~\cite{cassidy_coldPs:08}. The main features of the laser are briefly summarized hereafter. A pulse length of \SI{70}{\nano\second}, much longer than the spontaneous emission lifetime of the \otS{}--\tP{} transition (\SI{3.19}{\nano\second}), allows for multiple cooling cycles per pulse. The cavity is \SI{1}{\meter} long. The central wavelength is set by means of an intra-cavity Volume Bragg Grating (VBG)~\cite{Hemmer_vbg:2009}. Rotating the VBG finely tunes the fundamental wavelength with an absolute accuracy of \SI{10}{\pico\meter}. Two LBO (lithium triborate) crystals and two BBO (beta barium borate) crystals are used to generate \SI{2.3}{\milli\joule} at the third harmonic. At \SI{243}{\nano\meter}, the measured root-mean-square (rms) spectral bandwidth is $\sigma_{243}=\SI{101\pm{3}}{GHz}$. 
The laser~\cite{alex} was specifically designed to deliver an irradiance of \SI{100}{\kilo\watt\per\centi\meter\squared} when focusing \SI{0.7}{\milli\joule} on an area of \SI{10}{\milli\meter\squared}. As a result, the power on the \SI{20}{\mega\hertz} rms resonance transition line width amounts to $\text{\SI{100}{\kilo\watt\per\centi\meter\squared}}\times\text{\SI{20}{\mega\hertz}}/\text{\SI{101}{GHz}} =$ \SI{20}{\watt\per\square\centi\meter}, much higher than the saturation intensity of the \otS{}--\tP{} transition of \SI{0.45}{\watt\per\square\centi\meter}~\cite{zimmer_cooling:21}. The laser fluence fills in the spectral gaps in the laser bandwidth~\cite{cassidy_ryd:12} and the population in the excited state is saturated within a \SI{360\pm15}{\giga\hertz} large spectral bandwidth. 
It should be noted that in these conditions less than \SI{1.4}{\percent} of the atoms are photo-ionized \cite{zimmer_cooling:21}. 
The transverse Doppler profile was probed by fine-tuning the wavelength of a \SI{1.5}{ns} long \SI{205}{\nano\meter} pulse with a rms spectral bandwidth of $\sigma_{205} = $\SI{179\pm9}{GHz} or \SI{25\pm1}{pm}, populating the \ttP{} states. A \SI{4}{ns} long \SI{1064}{\nano\meter} pulse synchronized with the \SI{205}{\nano\meter} pulse induces the photo-ionization of the excited states \cite{aegis_neq3:16}.\\
\indent The Pockels cell of the alexandrite laser cavity is connected to a high-voltage electronic switch, which is opening and closing the cavity with nanosecond precision to generate a Q-switched pulse featuring a controllable sharp falling edge.  Consequently, the laser emission can be suppressed imminent to the arrival of the \SI{205}{\nano\meter} pulse probing the velocity profile, avoiding a temporal overlap of the cooling and probing laser pulses. The \SI{205}{\nano\meter} pulse interacts with Ps about \SI{12}{\nano\second} after the \SI{243}{\nano\meter} pulse has subsided. 
This guarantees that all transiently excited Ps atoms have spontaneously decayed to the ground state before probing the velocity distribution of the cloud. The synchronization of the lasers with the \pos{} implantation time is achieved with nanosecond time precision by utilizing the ARTIQ/Sinara framework \cite{aegis_sinara:22} and a LabVIEW-based distributed control system developed at \aegis{} \cite{aegis_talos:23}. 
The \SI{243}{\nano\meter} laser beam is co-propagating with the \SI{205}{\nano\meter} (see Fig.~\ref{fig:exp_scheme}) and retro-reflected by a dichroic mirror transmitting the \SI{205}{\nano\meter} light. All laser beams are linearly polarized. 

\begin{figure}[htp]
	\centering
	\includegraphics[width=\textwidth]{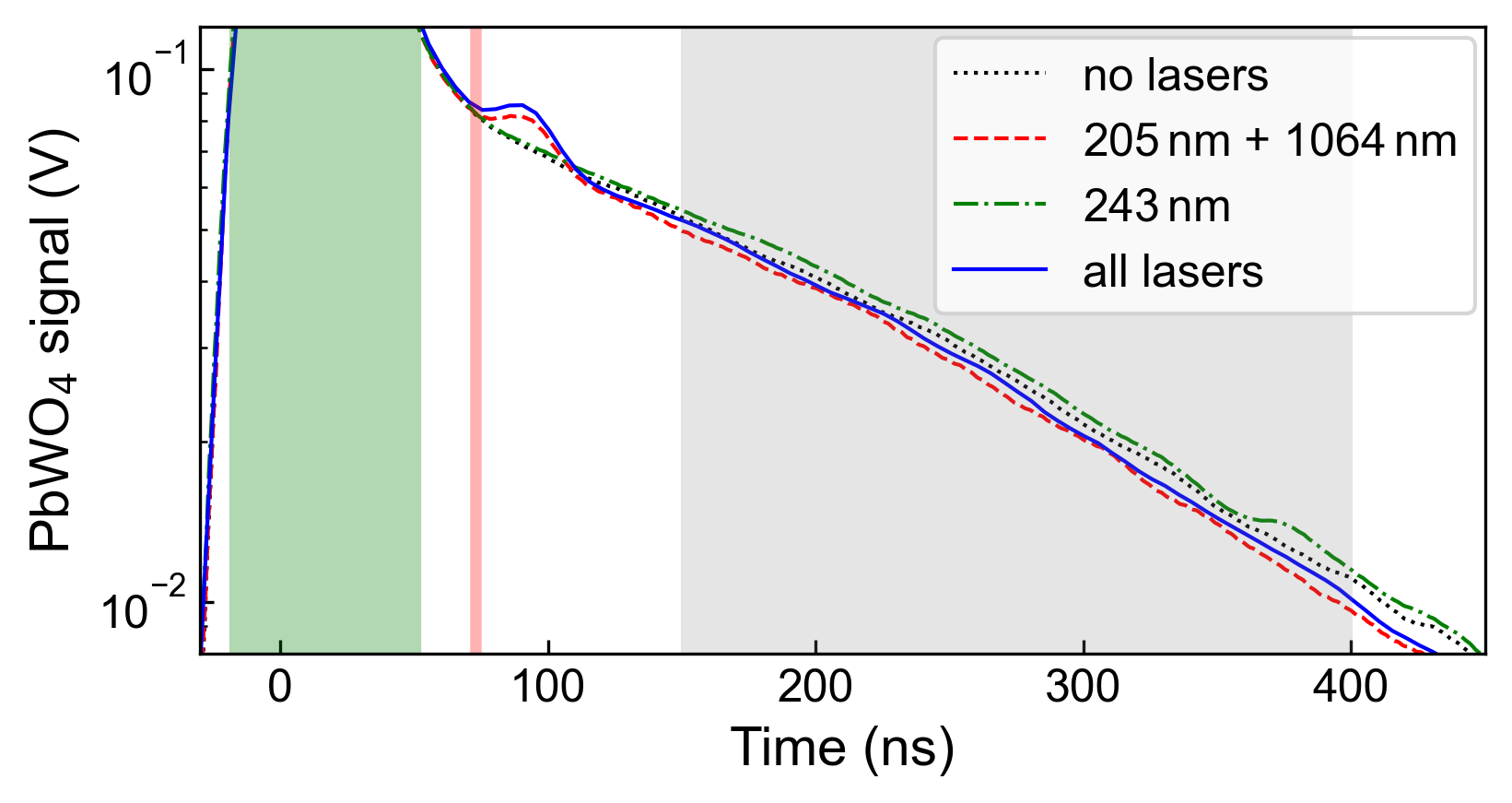}
	\caption{SSPALS spectra of Ps in vacuum without lasers (black dotted curve), \SI{205}{\nano\meter}+\SI{1064}{\nano\meter} lasers (red dashed curve), \SI{243}{\nano\meter} laser only (green dash-dotted curve), and all three lasers \SI{243}{\nano\meter}+\SI{205}{\nano\meter}+\SI{1064}{\nano\meter} (blue solid curve). The \SI{243}{\nano\meter} laser is firing during the time window from \SIrange{-20}{50}{ns} (green band), while the \SI{205}{\nano\meter}+\SI{1064}{\nano\meter} (red vertical line) are injected \SI{75}{\nano\second} after \pos{} implantation time ($t= \text{\SI{0}{\nano\second}}$). Each curve is an average of 90 individual spectra. For analysis, the window between \SI{150}{ns} and \SI{400}{ns} (light grey area) was used.}
	\label{fig:SSPALS}
\end{figure}

The time distribution of the gamma radiation resulting from Ps annihilation, the so-called Single-Shot Positron Annihilation Lifetime Spectroscopy (SSPALS)~\cite{cassidy_apl:06} spectrum, is acquired in different laser configurations. As illustrated in Fig.~\ref{fig:SSPALS}, the configurations ``no~lasers", ``\SI{205}{\nano\meter}+\SI{1064}{\nano\meter}", ``\SI{243}{\nano\meter} only", and ``\SI{243}{\nano\meter}+\SI{205}{\nano\meter}+\SI{1064}{\nano\meter}" are used. A $25\times25\,$mm $\mathrm{PbWO_4}$ scintillator placed \SI{40}{\milli\meter} above the \pos{}/Ps converter and coupled to a Hamamatsu R11265-100 photo-multiplier tube (PMT R11265U-200) connected to an oscilloscope (Teledyne LeCroy HDO4104A), is used to acquire the spectra at a rate of \SI{2.5}{GS\per\second}~\cite{aegis_neq3:16}. \\
\indent The long tail in the SSPALS spectrum measured without lasers (black dotted curve), extending from \SI{100}{\nano\second} to \SI{450}{\nano\second} in Fig.~\ref{fig:SSPALS}, reflects the \SI{142}{\nano\second} lifetime of \otS{} Ps in vacuum. Firing the \SI{243}{\nano\meter} laser only, a large fraction of Ps is excited to the \tP{} level, where the  annihilation lifetime is much longer than in the ground state. Consequently, the annihilation rate at later times increases 
(green dash-dotted line in Fig.~\ref{fig:SSPALS}) due to an increase of the number of annihilating atoms in the ground state. By sending only the \SI{205}{\nano\meter}+\SI{1064}{\nano\meter} pulses, a fraction of the atoms is selectively photo-ionized. This leads to an immediate small increase in the gamma emission rate due to annihilation of the photo-dissociated \pos{} hitting the nearby chamber walls (small bump at \SI{90}{\nano\second} in Fig.~\ref{fig:SSPALS}), followed by a reduction of the number of ground state Ps annihilating with \SI{142}{\nano\second} lifetime (red dashed curve). The interaction of the Ps cloud with all three lasers induces a combined effect (blue solid curve in Fig.~\ref{fig:SSPALS}).

In order to study the effects caused by the different laser configurations, S-parameters are constructed as:
\begin{equation}
S = \frac{f_{\text{ON}} - f_{\text{OFF}}}{f_{\text{OFF}}} \:, 
\label{Spercent}
\end{equation}
where $f_{\text{ON}}$ and $f_{\text{OFF}}$ denote the integrated SSPALS spectra in the time window between \SI{150}{\nano\second} and \SI{400}{\nano\second}. ON refers to one or several of lasers interacting with the Ps cloud, and  OFF to no laser interaction. In the following, we will refer to $S_{\text{205+1064}}$ when only the probing lasers are present, $S_{\text{243}}$ when only the cooling laser is present. We further define $S_{\text{cool}}$ as the difference between $S_{\text{243+205+1064}}$ (when all three lasers are present) and $S_{\text{243}}$:
\begin{equation}
S_{\text{cool}} = S_{\text{243+205+1064}}-S_{\text{243}}  
\label{Scool}
\end{equation}
$S_{\text{cool}}$ reflects the number of photo-ionized Ps atoms for a given probing laser detuning after interaction with the cooling laser normalized to the number of Ps atoms annihilating in the absence of any laser.
High statistics on the S-parameter values is achieved by averaging over sets of spectra acquired consecutively in all the above-mentioned laser configurations. A detrending procedure is applied \cite{aegis_meta:18} to correct for slow changes in the amount of Ps produced over time, caused by moderator aging during the long measurement period.
A trend function is built by applying Gaussian radial basis regression \cite{scikit-learn} to the $f_{\text{OFF}}$ data. Subsequently, S-parameters are calculated by evaluating the trend function at exactly the time at which the SSPALS spectra with laser(s) are acquired.

\begin{figure}[htbp]
	\centering
	\includegraphics[trim=5 5 5 5,clip,width=0.49\textwidth]{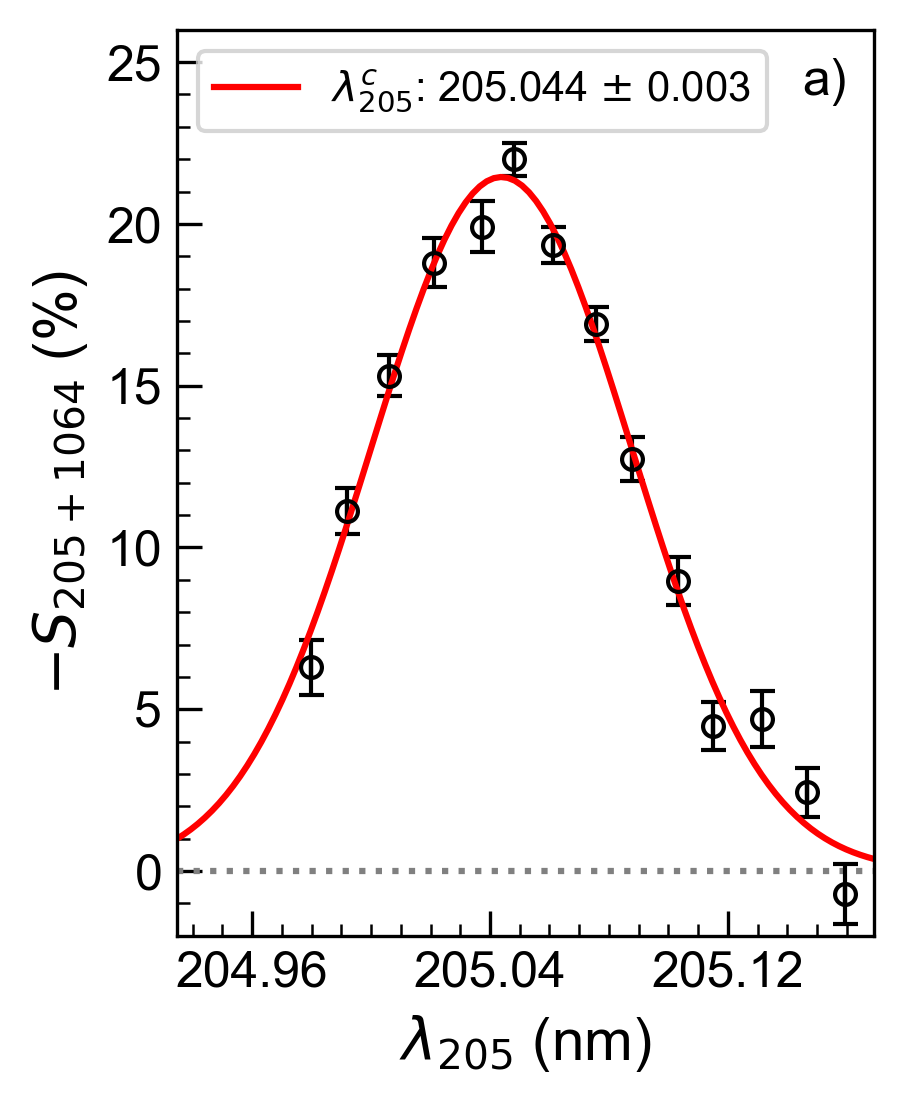}
    \includegraphics[trim=5 5 5 5,clip,width=0.49\textwidth]{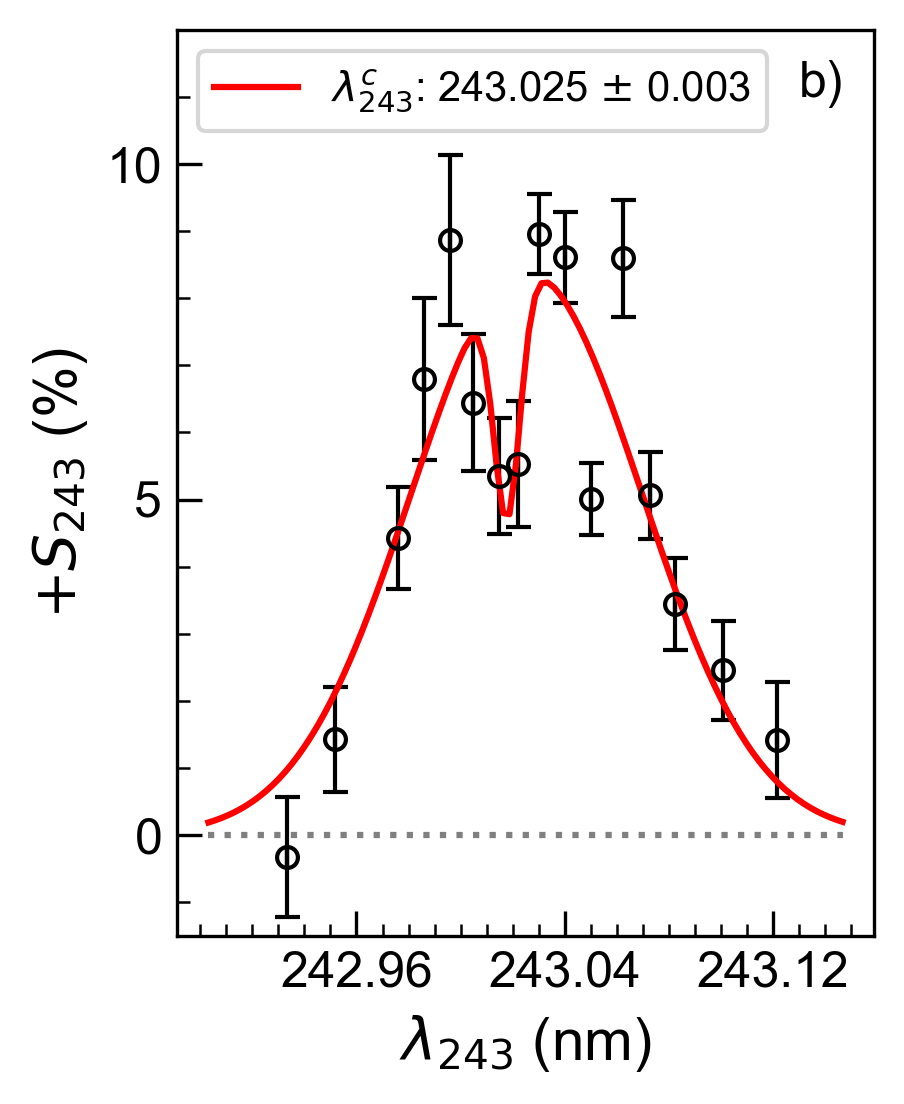}
	\caption{Ps velocity distribution measured by SSPALS. \\
    a) Transverse Doppler profile measured by two-photon resonant ionization (205 and \SI{1064}{\nano\meter} lasers) by means of the $S_{\text{205+1064}}$ parameter. A Gaussian fit yields a rms-width of \SI{44\pm1}{pm}, which translates to a Ps rms-velocity of \SI{5.3\pm0.2e4}{\meter\per\second} after de-convolution of the laser bandwidth.\\
    b) Velocity-resolved increase in the number of ground state Ps atoms, induced by the \SI{243}{\nano\meter} transitory excitation to the \tP{} level, represented by the $S_{\text{243}}$ parameter. At resonance, the Lamb dip is clearly visible. A 2-Gaussian fit yields a rms-width of the engulfing Gaussian of \SI{44\pm3}{pm}, which corresponds to a Ps rms-velocity of \SI{4.9\pm0.4e4}{\meter\per\second}.}
	\label{fig:doppler_ref}
\end{figure}

First, the Ps velocity distribution without laser cooling was measured by scanning the $\lambda_{205}$ detuning with the \SI{205}{\nano\meter} and \SI{1064}{\nano\meter} laser beams grazing the surface of the \pos{}/Ps converter. In this experiment, the laser pulses were fired \SI{50}{\nano\second} after the positron implantation time. The transverse Doppler profile, shown in Fig.~\ref{fig:doppler_ref}~a), is fitted with a Gaussian function, yielding a rms-width of \SI{44\pm1}{pm}. This width corresponds to a Ps rms-velocity of $\SI{5.3\pm0.2e4}{\meter\per\second}$ after de-convolution of the probing laser bandwidth (\SI{25\pm1}{pm}). The resulting line is centered at $\lambda^{c}_{205} = \text{\SI{205.044\pm0.003}{\nano\meter}}$, which is compatible with the theoretical value~\cite{aegis_neq3:16}.

Secondly, we performed Saturated Absorption Spectroscopy on the \otS{}--\tP{} transition~\cite{cassidy_hyp:12} to determine the center of the line and the effect of the cooling pulse on the SSPALS spectrum. 
In Fig.~\ref{fig:doppler_ref}~b), SSPALS spectra are recorded as function of the $\lambda_{243}$ detuning with the cooling laser pulse synchronized with the \pos{} implantation time (see Fig.~\ref{fig:SSPALS}). 
It is worth noting that the resulting S-values are now positive, in contrast to what was observed in the two-photon resonant ionization experiment. To the best of our knowledge, such an increase in the number of ground state Ps atoms caused by a transitory laser excitation to the \tP{} level has never been observed and can be classified as a laser-induced, spectrally tunable preservation of Ps atoms. This effect has to be distinguished from the lifetime enhancement of Ps atoms excited to Rydberg-states~\cite{cassidy_ryd:12}.

The observed line shape shows a Lamb dip.  
A 2-Gaussian model is fitted to the data, following the modeling suggested in Ref.~\cite{cassidy_hyp:12}. The transition line is centered at $\lambda^{c}_{243}=\text{\SI{243.025\pm0.003}{\nano\meter}}$, which is in agreement with previous measurements~\cite{cassidy_hyp:12}.
The engulfing Gaussian features a rms-width of $\SI{44\pm3}{pm}$, which corresponds to a Ps rms-velocity of \SI{4.9\pm0.4e4}{\meter\per\second} after de-convolution of the $\sigma_{243}$ cooling laser bandwidth (\SI{20\pm1}{pm}). Hence, this measurement allows to confirm the transverse 1D-velocity profile of our original Ps cloud. 
Furthermore, the measured width of the Lamb dip is compatible with previous observations~\cite{cassidy_hyp:12}  and demonstrates the saturation of the \otS{}--\tP{} transition by the laser.

With this understanding of the individual laser interactions with the Ps cloud, we then performed experiments combining the \SI{243}{\nano\meter} cooling laser and the \SI{205}{\nano\meter}+\SI{1064}{\nano\meter} probing lasers. The cooling laser remains in the same spatial position,
whereas the probing laser beam is moved to a position at a distance of \SI{7}{\milli\meter} from the converter surface (see Fig.~\ref{fig:exp_scheme}). 
Accordingly, the probing laser pulses are delayed by \SI{75}{\nano\second} with respect to the positron implantation time, as indicated by the red vertical line in Fig.~\ref{fig:SSPALS}. This delay corresponds to the time taken by the atoms in the peak-velocity component of the axial velocity distribution to reach the position of the laser beam. 

\begin{figure}[htp]
	\centering
	\includegraphics[trim=5 5 5 5,clip,width=\textwidth]{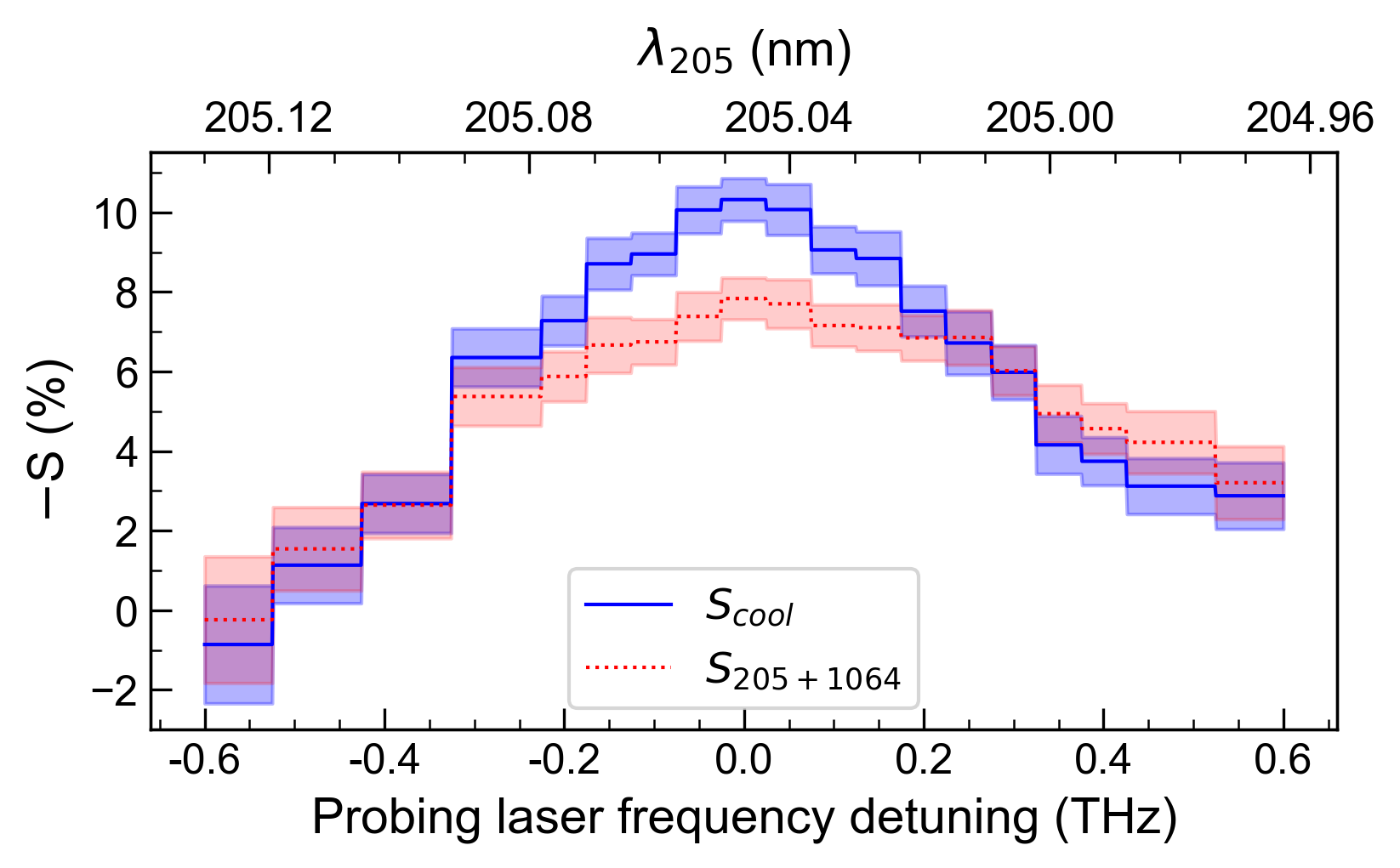}
	\caption{One-dimensional transverse Doppler profiles of the Ps cloud with (solid curve), and without (dotted curve) interaction with the \SI{243}{\nano\meter} cooling laser beam at a fixed frequency detuning of \SI{-200}{GHz}. The semi-transparent bands represent the statistical measurement error (one standard deviation of the average).}
	\label{fig:Cooling_vs._nocooling.jpg}
\end{figure}

To characterize the change in the Ps velocity distribution induced by the cooling laser, the detuning of the \SI{243}{\nano\meter} laser is set to \SI{-200}{\giga\hertz} (corresponding to $\lambda_{243}=\SI{243.061}{nm}$) and a photo-ionization Doppler scan is performed with the \SI{205}{\nano\meter}+\SI{1064}{\nano\meter} lasers. The $S_{\text{cool}}$ parameter measured as a function of $\lambda_{205}$ is shown in Fig.~\ref{fig:Cooling_vs._nocooling.jpg}. The curve is compared to the $S_{\text{205+1064}}$ distribution measured in the same configuration (\SI{75}{\nano\second} delay and \SI{7}{\milli\meter} away from the \pos{}/Ps converter), but without prior interaction with the cooling laser. Both of the one-dimensional transverse Doppler profiles were obtained by applying a moving average to the $\sim\,$350 single S-values with a square window (\SI{350}{GHz} in width).

The one-dimensional transverse Doppler profile obtained in the presence of the \SI{243}{\nano\meter} cooling laser is narrower than the one measured without it. The asymmetry of the two profiles is caused by a slight increase in the pulse energy of the \SI{205}{nm} probing laser toward blue-detuned wavelengths. 
As a first-order estimate of the cooling efficiency, we use a simple Gaussian fit on each of the two distributions to quantify the change in the velocity profile. With prior cooling, we find a rms-width of \SI{269\pm1}{GHz}, in contrast to \SI{330\pm2}{GHz} without cooling. 
After de-convolution of the standard deviation of the moving average window ($\SI{350}{GHz} / \sqrt{12}$) and of the $\sigma_{205}$ laser bandwidth (\SI{179}{GHz}), the Ps rms-velocities corresponding to these widths are $\SI{5.4\pm0.2e4}{\meter\per\second}$, associated with a temperature of $\SI{380\pm20}{\kelvin}$, and  $\SI{3.7\pm0.2e4}{\meter\per\second}$ associated with $\SI{170\pm20}{\kelvin}$, respectively.
The obtained rms-velocity in the absence of the cooling laser is in agreement with the result reported in Fig.~\ref{fig:doppler_ref}~a) when the 205 and \SI{1064}{\nano\meter} lasers were grazing the target. 
The interaction with the \SI{70}{\nano\second} long \SI{243}{\nano\meter} laser pulse reduces the Ps rms-velocity by \SI{1.7\pm0.3e4}{\meter\per\second}, corresponding to a temperature reduction of $\Delta T = \SI{210\pm30}{\kelvin}$. The systematic error associated with the arbitrary choice of a Gaussian fitting model is estimated to be $\pm\SI{10}{\kelvin}$.

Given the high optical intensity of the \SI{243}{\nano\meter} laser, the average time for all addressed Ps atoms to undergo a single cooling cycle is \SI{6.38}{\nano\second}~\cite{cassidy_coldPs:08}. Consequently, with a \SI{70}{\nano\second}-long Ps-laser interaction, a maximum of 11 cycles is possible. 
Since the recoil velocity for a single \otS{}--\tP{} transition of Ps is $v_{\text{recoil}}=\SI{1.5e3}{\meter\per\second}$~\cite{zimmer_cooling:21}, the velocity reduction can reach $11\times v_{\text{recoil}}=\SI{1.65e4}{\meter\per\second}$, corresponding to a temperature reduction of about \SI{200}{K}, in agreement with our measurements.

\begin{figure}[htp]
	\centering
	\includegraphics[trim=5 5 5 5,clip,width=\textwidth]{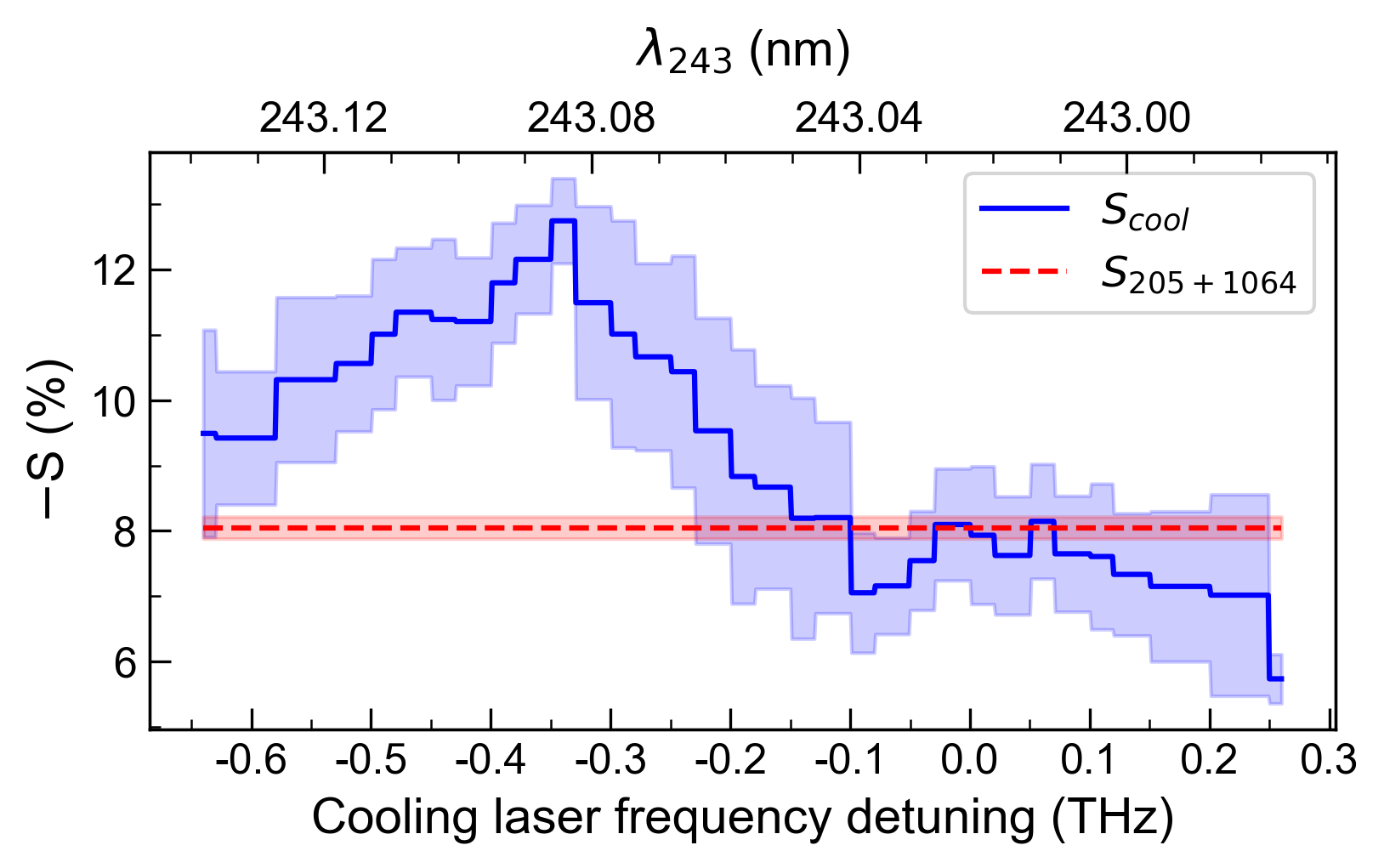}
	\caption{Number of Ps atoms with a velocity within the bandwidth of the \SI{205}{\nano\meter} probing laser, which is kept at resonance, as a function of the cooling laser frequency detuning, normalized to the number of Ps atoms in the absence of all lasers. The dashed horizontal line represents the reference population of Ps atoms in this velocity range with the cooling laser off. The amount of Ps atoms in the center increases by up to \SI{58\pm9}{\percent} at a cooling laser frequency detuning of \SI{-350}{\giga\hertz}. The semi-transparent bands represent the statistical uncertainties (one standard deviation of the average).}
	\label{fig:alex_scan}
\end{figure}

To evaluate the maximum fraction of fast Ps atoms that can be pushed toward null velocity via recoil effect, the cooling laser detuning was scanned from \SIrange{-0.65}{0.25}{THz} while the \SI{205}{\nano\meter} laser remained at resonance. 
The result of this scan is shown in Fig.~\ref{fig:alex_scan}. The horizontal dashed line is the signal measured when only the probing laser interacts with the Ps cloud ($S_{\text{205+1064}}$), yielding $-S=\SI{8.0\pm0.2}{\percent}$ as the reference for the population near resonance. The blue curve is the $S_{\text{cool}}$ parameter as defined in Eq.~\ref{Scool}. The curve was obtained by applying a moving average with a window size of \SI{200}{GHz}.
For a given $\lambda_{243}$ detuning, the difference between $S_{\text{cool}}$ and $S_{\text{205+1064}}$ corresponds to the fraction of the Ps population cooled within the bandwidth of the \SI{205}{\nano\meter} laser. 
A maximum increase of \SI{58\pm9}{\percent} compared to the reference signal is reached at a detuning of \SI{-350}{\giga\hertz}. 

In conclusion, we have experimentally demonstrated laser cooling of a large fraction of a thermal Ps cloud in a magnetic and electric field-free environment. A temperature decrease from \SI{380\pm20}{\kelvin} to \SI{170\pm20}{\kelvin} was observed. Our study also gives an in-depth understanding of the different laser--Ps interactions and their manifestation in the SSPALS spectra recorded during the experiment. In particular, we observed an increase in the number of atoms in the ground state after Ps has been transiently excited to the \tP{} states. As a result, our cooling method has the unique feature of delaying the annihilation, which allows to preserve a large number of Ps atoms while cooling the ensemble. Our results can be improved by adding a second cooling stage with a narrower spectral bandwidth set to a detuning closer to resonance, or by coherent laser cooling~\cite{Corder15_CoherentCooling,Bartolotta18}. In addition, this technique can be extended to two and three-dimensional laser cooling. The cooling of Ps opens the door to an entirely new range of important fundamental studies with cold Ps beams, including 
precision spectroscopy, Bose-Einstein condensation of Ps, and tests of the Equivalence Principle with a purely leptonic matter-antimatter system. 

\subsection{Acknowledgement}

The authors are grateful to P. Yzombard, C. Zimmer and O. Khalidova for early contributions to this activity, to L. Cabaret for the development of the \otS{}-\tP{} laser and to Dr. S. Cialdi for the original development of the \otS{}-\ttP{} laser. This work was supported by 
the ATTRACT program under grant agreement EU8-ATTPRJ (project O--Possum II);
Istituto Nazionale di Fisica Nucleare; 
European Union's Horizon 2020 research and innovation programme under the Marie Sklodowska-Curie Grant Agreement No. 754496, FELLINI and No. 748826, ANGRAM; 
the CERN Fellowship programme and the CERN Doctoral student programme; 
%
the EPSRC of UK under grant number EP/X014851/1;
%
Research Council of Norway under Grant Agreement No. 303337 and NorCC;
%
NTNU doctoral program;
the Research University – Excellence Initiative of Warsaw University of Technology via the strategic funds of the Priority Research Centre of High Energy Physics and Experimental Techniques;
the IDUB POSTDOC programme;
the Polish National Science Centre under agreements no. 2022/45/B/ST2/02029, and no. 2022/46/E/ST2/00255, and by the Polish Ministry of Education and Science under agreement no. 2022/WK/06;
%
Marie Sklodowska-Curie Innovative Training Network Fellowship of the European Commission's Horizon 2020 Programme (No. 721559 AVA); 
Wolfgang Gentner Programme of the German Federal Ministry of Education and Research (grant no. 13E18CHA); 
European Research Council under the European Unions Seventh Framework Program FP7/2007-2013 (Grants Nos. 291242 and 277762); 
the European Social Fund within the framework of realizing the project, in support of intersectoral mobility and quality enhancement of research teams at Czech Technical University in Prague (Grant No. CZ.1.07/2.3.00/30.0034);
%
%
%
%

\bibliographystyle{apsrev}
\bibliography{aegis_biblio}

\end{document}